# HEART-Watch: A multimodal physiological dataset from a Google Pixel Watch across different physical states

Jathushan Kaetheeswaran

Institute of Biomedical Engineering, University of Toronto, Canada and KITE Research Institute, University Health Network, Canada

Boyi Ma

Institute of Biomedical Engineering, University of Toronto, Canada and KITE Research Institute, University Health Network, Canada

Ali Abedi

KITE Research Institute, University Health Network, Canada

Shehroz S. Khan*

College of Engineering and Technology, American University of the Middle East, Egaila, 54200, Kuwait

Milad Lankarany*

Institute of Biomedical Engineering, University of Toronto, Canada and KITE Research Institute, University Health Network, Canada

* These authors contributed equally to this work

Consumer-grade smartwatches offer a new option for personalized health monitoring for general consumers, as cardiovascular diseases continue to prevail as the leading cause of global mortality. The development and validation of reliable cardiovascular monitoring algorithms for these consumer-grade devices requires realistic biosignal data from diverse sets of participants. However, the availability of public consumer-grade smartwatch datasets with synchronized cardiovascular biosignals remains limited, and existing datasets often lack rich demographic diversity in their participant cohorts, potentially leading to biased algorithm development. This paper presents HEART-Watch, a multimodal physiological dataset of synchronized wrist-worn Google Pixel Watch electrocardiogram (ECG), photoplethysmography, and accelerometer signals from a diverse cohort of 40 healthy adults across three physical states – sitting, standing and walking – alongside reference chest ECG. Intermittent upper arm blood pressure measurements and concurrent biosignals were collected as an additional biomarker for future research. The motivation, methodology, and initial analyses of results are presented. HEART-Watch is intended to support the development and benchmarking of robust cardiovascular algorithms on consumer-grade smartwatches across diverse populations.

## AUTHOR SUMMARY

Consumer-grade smartwatches are driving a global shift towards continuous and accessible healthcare for the general population. However, developing robust and equitable cardiovascular monitoring algorithms for these devices requires high-quality datasets for training and validation. Many consumer-grade smartwatches restrict access to raw sensor data, limiting researchers' ability to obtain high-quality biosignals directly from the devices. Additionally, participant demographics in existing datasets often do not reflect the diversity present in real-world conditions, frequently focusing on young adults from a single ethnic background. Here, we establish the HEART-Watch dataset to address these challenges by collecting synchronized biosignals directly from a Google Pixel Watch across sitting, standing, and walking conditions in a diverse cohort of 40 participants. This work provides the research community with a realistic and inclusive resource to develop and evaluate cardiovascular monitoring algorithms for consumer-grade devices across a broad range of ages, body types, and backgrounds.

## 1 INTRODUCTION

Remote cardiovascular monitoring has advanced significantly over the past decade with the release of novel consumer-grade smartwatches and artificial intelligence (AI) algorithms. Given that cardiovascular diseases (CVDs) continue to be the primary cause of mortality globally [1,2], continuous monitoring with commercial smartwatches can be leveraged to identify risk factors, promote healthier lifestyles, and reduce burdens on healthcare systems by encouraging self-monitoring habits [3–5]. Modern smartwatches integrate electrocardiogram (ECG), photoplethysmography (PPG), and accelerometer (ACC) sensors to noninvasively acquire physiological signals for cardiovascular analysis [6,7]. The rapid increase in demand for accessible health monitoring tools has propelled the smartwatch industry to an estimated global market value of $39 billion USD by 2034 [8], showcasing the widespread integration of these tools into the lives of the general public. Clinicians are also beginning to incorporate data from wearables into clinical consultations, such as for atrial fibrillation (AF) screenings [9,10].

Despite their increasing prevalence in both personal health monitoring and clinical contexts [11–13], consumer-grade smartwatches are not always classified as medical devices, thereby exempting them from rigorous and transparent validation studies [6,7,14–16]. For instance, heart rate (HR) estimation on WearOS relies on PPG and ACC sensors and is exempt from premarket approval by the US Food and Drug Administration (FDA) as it is considered a low-risk "wellness" feature [6,7]. This regulatory context raises concerns about the validity and reliability of derived cardiovascular metrics. Smartwatches that incorporate medical-grade sensors or functionality, such as ECG, may receive expedited US FDA clearance by demonstrating substantial equivalence to existing predicate devices – legally marketed devices previously cleared with the same intended use and technological characteristics – rather than undergoing more rigorous premarket approval processes [6,17]. For example, the Apple Watch received US FDA clearance for an AI-based hypertension-notification feature by demonstrating substantial equivalence to Viz HCM, an algorithm operating on 12-lead ECG, despite the Apple feature operating on wrist-based PPG [18,19]. This example illustrates that regulatory clearance does not necessarily require substantial equivalence using consumer-grade smartwatch sensors and data; rather, it may be granted by demonstrating similar intended use and performance.

Robust validation of consumer-grade smartwatches is critical given the limitations of the embedded sensors on wearable devices. The ECG, PPG, and ACC signals collected by wrist-worn devices are vulnerable to quality issues, particularly during physical activity, such as motion artifacts, signal noise and sensor displacements, reducing their effectiveness for individuals across regular daily activities [6,20–22]. Modern smartwatches are often deployed with AI algorithms that are trained and/or validated on smartwatch datasets [23–25], indicating that the quality of these datasets dictate real-world performance. To contextualize this, a non-exhaustive summary of publicly available, non-synthetic physiological datasets from consumer-grade smartwatches is presented in Table 1. Datasets were identified through online searches of research articles using terms such as "smartwatch dataset", "wearable dataset", and "consumer-grade wearable dataset", supplemented by citation tracking. Additional datasets were identified from studies on smartwatch-based cardiovascular algorithms in which data collection was performed but not the primary focus of the work. Datasets were excluded if (i) raw signals or aggregate measures were not provided, (ii) the focus of the work was not cardiovascular analysis for awake, healthy adults, (iii) data collection involved invasive, custom or prototype devices, or (iv) authors did not provide data availability statements or clearly stated that data was not available to researchers (e.g., the Apple Heart Study [24] and Fitbit Heart Study [23]).

Table 1. Non-exhaustive literature review of available consumer-grade smartwatch datasets for non-synthetic physiological data from awake, healthy adults.

| **Reference** | **Number of participants (F/M)** | **Mean age (standard deviation) years** | **Racial categories reported (n)** | **Height and/or weight recorded?** | **Smartwatch models** | **Data modalities** | **Duration** | **Physical states** | **Data availability** |
|---|---|---|---|---|---|---|---|---|---|
| Bent et al. 2020 [21,26] | 53 (32/21) | 25.6 (—) | 4 | ✓ | Apple Watch 4, Fitbit Charge 2, Garmin Vivosmart 3, Xiaomi Miband 3, Empatica E4, Biovotion Everion | Chest HR, HR | 1-5 minutes | Sitting, Paced deep breathing, Walking, Typing | Data use agreement |
| Reece et al. 2021 [27] | 23 (10/13) | 23.6 (3.2) | — | ✓ | Apple Watch 4, Garmin Forerunner 735XT | Chest HR, HR | 57 minutes | Sitting, Everyday activities, Walking, Jogging, Running, Cycling | Upon request |
| Sarhaddi et al. 2022 [28] | 28 (14/14) | 32.5 (6.6) | — | — | Samsung Gear Sport | Chest ECG, PPG | 24 hours | Free-living | Data use agreement |
| SmartHeartWatch 2022 [29] | 18 (11/7) | 56.0 (20.0) | — | — | Apple Watch 7, | ECG | 30-50 seconds | Resting, Recovery | — |

| Reference | Number of participants (F/M) | Mean age (standard deviation) years | Racial categories reported (n) | Height and/or weight recorded? | Smartwatch models | Data modalities | Duration | Physical states | Data availability |
|---|---|---|---|---|---|---|---|---|---|
| | | | | | Withings ScanWatch | | | | |
| Jiang et al. 2023 [30] | 49 (18/31) | 63.1 (8.6) | 2 | — | Apple Watch Series 7, Garmin Fenix 6 Pro, Garmin Venu 2s, Withings ScanWatch | HR, SpO2 | 6 measures | Resting | Open access |
| SMART Start 2023 [29] | 28 (28/0) | 34.0 (4.0) years | — | — | Withings ScanWatch | ECG | 30-50 seconds | Resting | — |
| Biswas and Ashili 2023 [31] | 1 (0/1) | 48.0 (—) | 1 | — | Fitbit version 2 | HR | 1 month | Free-living | Subscription |
| Jaiswal et al. 2024 [32] | 32 (—) | — | — | — | Samsung Galaxy Watch 4 | Chest ECG, PPG | 3 minutes | Sitting, Standing, Walking | Upon request |
| Baigutanova et al. 2025 [33,34] | 49 (25/24) | 28.4 (5.9) | — | ✓ | Samsung Galaxy Active 2 | ACC, HR, PPG | 4 weeks | Free-living | Open access |
| GalaxyPPG 2025 [35,36] | 24 (12/12) | 23.3 (2.0) | 1 | — | Samsung Galaxy Watch 5 | Chest ECG, ACC, HR, PPG | > 1 hour | Social stress activities, Everyday activities | Open access |

| Reference | Number of participants (F/M) | Mean age (standard deviation) years | Racial categories reported (n) | Height and/or weight recorded? | Smartwatch models | Data modalities | Duration | Physical states | Data availability |
|---|---|---|---|---|---|---|---|---|---|
| Hung et al. 2025 [37,38] | 25 (16/9) | 25.8 (1.9) | — | ✓ | Fitbit Charge 5 | Chest ECG, PPG | 3-5 minutes | Resting, Cycling | Open access |
| Ravanelli et al. 2025 [39,40] | 26 (14/12) | 30.1 (14.2) | — | ✓ | Fitbit Charge 5, Bangle.js2 | Chest HR, HR | 24 hours | Free-living | Open access |
| **HEART-Watch (this paper)** | **40 (23/17)** | **44.2 (21.0)** | **8** | ✓ | **Google Pixel Watch (2022)** | **Chest ECG, ACC, ECG, PPG, Cuff BP** | **4 minutes** | **Sitting, Standing, Walking** | **Data use agreement** |

Modalities should be considered smartwatch-derived unless indicated. A '—' indicates that the information was not readily reported.

One of the primary limitations of existing datasets is the availability and duration of smartwatch ECG recordings, with only two datasets, SmartHeartWatch and SMART Start, providing this modality [29]. Due to the restrictions on data exports from consumer-grade smartwatches, existing datasets containing smartwatch ECGs are restricted to short recordings at rest, typically 30 seconds, as these are often the default settings in traditional applications [41]. The lack of publicly available prolonged smartwatch ECGs is also likely due to technical challenges associated with wrist-based acquisition, which requires users to maintain constant contact with the opposite hand's index finger [42,43]. Despite these limitations, smartwatch ECGs provide greater clinical value, as physicians are more inclined to interpret ECG tracings than PPG signals, highlighting the importance of this modality in dataset design [10]. Similarly, consumer-grade devices generally make it challenging to access raw sensor data, so existing datasets often only provide aggregate measures such as HR and blood oxygen saturation (SpO2) [21,26,27,30,31,39,40]. While these aggregate measures enable downstream analyses, the underlying algorithms that map sensor data to these measures are typically proprietary and undisclosed, limiting reproducibility and transparency [44].

Demographic information is also consistently underreported in these datasets, with only four reporting any details regarding racial diversity [21,26,30,31,35,36] and only five providing height and weight information [21,26,27,33,34,39,40]. Many existing datasets do not explicitly recruit cohorts that reflect real-world heterogeneity, despite growing evidence that these factors significantly influence signal quality, derived features, and algorithmic bias. As a result, underreported and limited diversity in training and validation cohorts remains a persistent barrier to equitable performance for wearables, as models trained on non-representative data may learn biased relationships that fail to generalize to broader populations [6,37,45,46]. For example, wrist-based PPG has been shown to perform worse for HR estimation in individuals with darker skin tones (e.g., South Asian, Middle Eastern, or Black populations) during exercise compared to those with lighter skin tones [37]. Similarly, body height is known to be an important contributor to the time taken for PPG pulse arrival times at peripheral sites [15,47], which is critical for estimating metrics such as blood pressure (BP). Although many of these challenges are due to limitations of optical sensors, similar disparities should also be considered in metrics derived from wrist-based ECGs and multimodal approaches. Finally, as illustrated in Table 1, existing datasets are often heavily skewed towards young adult populations, with mean ages ranging from 23 to 34 years across most works. Older adults are rarely included in existing smartwatch datasets; for example, GalaxyPPG offers long-term, semi-naturalistic recordings, but only for young adults below the age of 30 [36]. The inclusion of older adults for smartwatch datasets is important to account for age-related physiological changes, such as lower skin perfusion and increased arterial stiffness [7,17,47,48], which can affect downstream algorithmic tasks. For instance, heart rate variability (HRV) estimation has been shown to be less accurate in individuals over 40 years of age [49].

Based on these limitations, a critical gap remains: open, multimodal datasets from consumer-grade smartwatches with rich participant demographics and recordings across physical states are scarce. To address this gap, this work presents HEART-Watch, a dataset of time-aligned ECG, PPG, and ACC signals from a consumer-grade smartwatch alongside reference chest ECG in a diverse cohort of healthy adults. Data were collected across sitting, standing, and walking physical states to enable controlled evaluations of algorithm robustness to common real-world signal contaminants, including motion artifacts and signal saturation, when extracting cardiovascular biomarkers such as HR and HRV [50]. Additionally, cuff BP measurements and simultaneous smartwatch ECG, PPG, and ACC recordings were acquired between physical states as an additional resource for multimodal cardiovascular analyses.

## 2 RESULTS

The preliminary validation of HEART-Watch is composed of (i) a visualization of synchronized signals, (ii) an assessment of signal sampling rates, (iii) an assessment of signal strength, (iv) statistical comparisons between chest and smartwatch ECG morphology, (v) an example of HR estimation techniques incorporating all smartwatch modalities, and (vi) an analysis of systolic and diastolic BP measurements.

### 2.1 Participants

Forty healthy adults (23 females, 17 males, age range = 19 – 75 years, age mean ± age standard deviation = 44.2 ± 21.0 years) were recruited to participate in this study from May – July 2025. Demographic information, including age, sex, height, weight, and race, was collected at the start of each session and is summarized in Table 2. The cohort supports age-based comparisons of smartwatch algorithms, with 40% (n = 16) of participants aged 55 years or older, and 52.5% (n = 21) below 40 years old. As shown in Fig 1, participants self-identified across eight racial groups, with the largest groups being East Asian (n = 12), White (n = 12), South Asian (n = 6) and Middle Eastern (n = 5). This racial diversity enables subgroup and fairness analyses during validation [45].

Table 2. Summary of HEART-Watch dataset characteristics, including participant demographics (age, height, weight and race) stratified by sex.

| **Characteristic** | **HEART-Watch Dataset** | |
|---|---|---|
| | **Female (n = 23)** | **Male (n = 17)** |
| Age, years | 41.87 ± 21.51 (19 – 72) | 47.23 ± 19.51 (21 – 75) |
| Height, meters | 1.63 ± 0.58 (1.53 – 1.72) | 1.79 ± 0.82 (1.65 – 1.94) |
| Weight, kilograms | 60.57 ± 11.84 (43 – 93) | 80.35 ± 12.47 (64 – 115) |
| BMI, $kg/m^2$ | 22.84 ± 4.27 (16.8 – 35) | 25.21 ± 3.92 (19.6–35.5) |

Values are reported as mean ± standard deviation (range) where applicable.

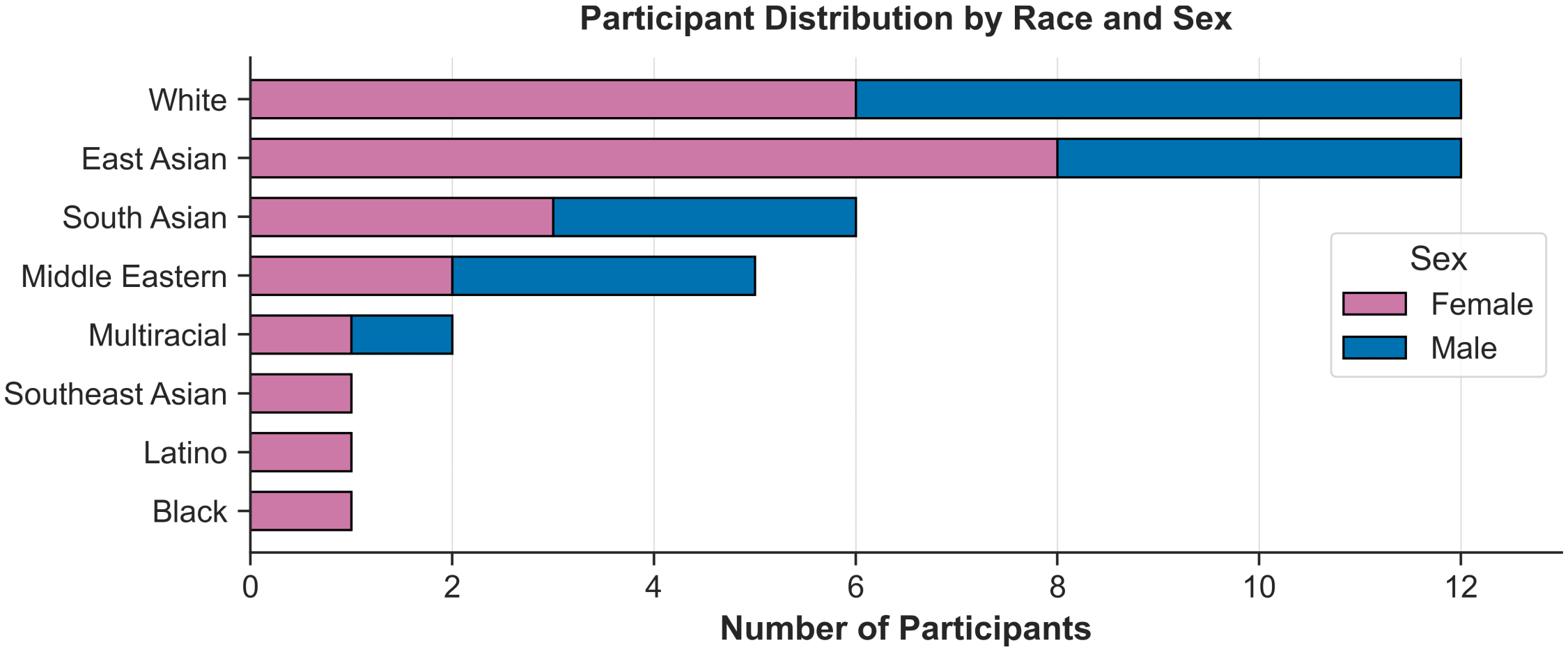

Fig 1. Distribution of participants' self-reported racial groups, stratified by sex.

Body-type diversity is also reflected in the height-weight distribution visualized in Fig 2. For example, among female participants with heights around 1.60 m, body weight ranged from 43 to 93 kg, highlighting substantial variability in body compositions within this subgroup. The broad range of participant heights (males = 1.65 – 1.94 m, females = 1.53 – 1.78 m) is relevant for PPG analyses, where body height has been recommended as a factor influencing HR estimation [15]. Future work could evaluate if similar associations exist for smartwatch ECG-derived metrics using this dataset.

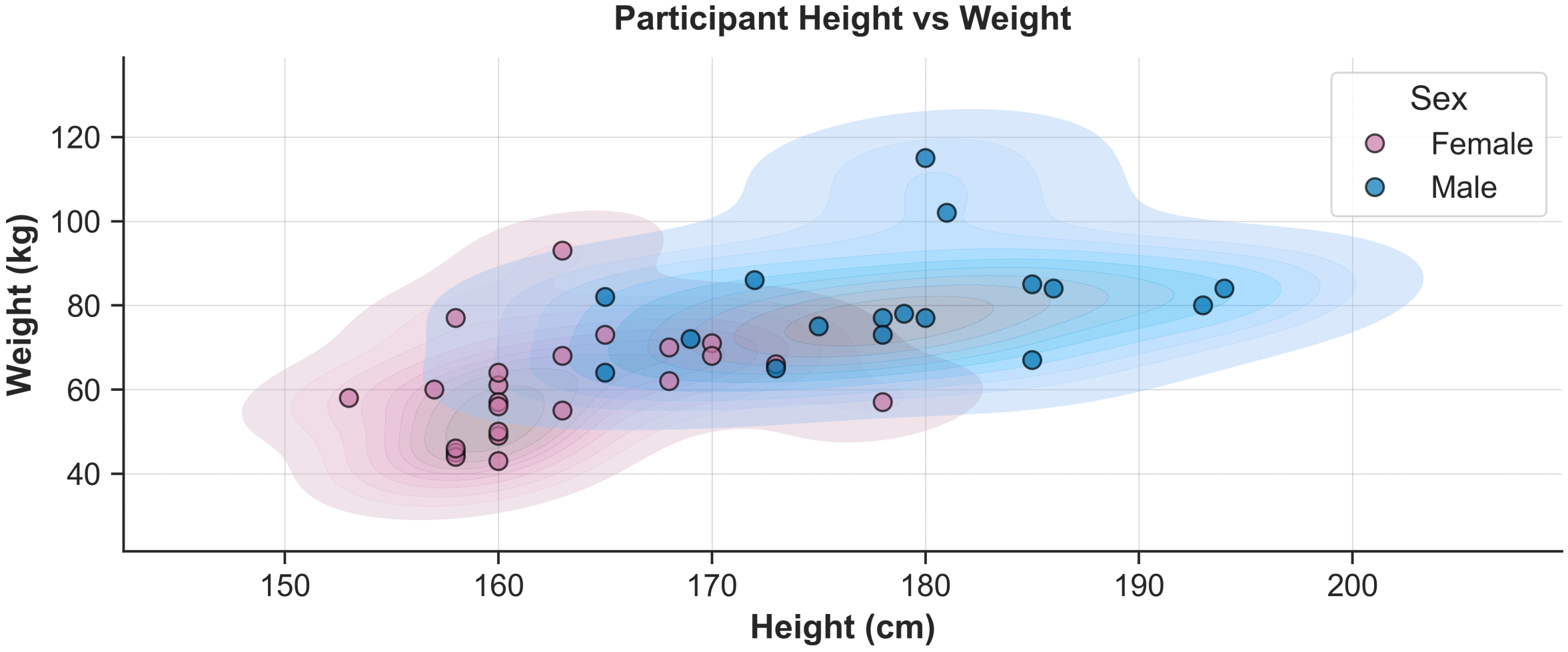

Fig 2. Distribution of participant height and weight shown using kernel density estimate with overlaid scatter plot, stratified by sex.

## 2.2 Visualization of Synchronized Signals

The synchronization of the smartwatch and chest signals is illustrated in Fig 3. The waveforms are presented after applying min-max normalization, mean removal, and filtering using Neurokit2, a common open-source Python toolbox for biosignal processing [51].

The R-wave peaks in both the chest and smartwatch ECG signals are closely aligned temporally across physical states, demonstrating that high-quality wrist-based ECG acquisition is feasible under non-static conditions. The PPG signal exhibits periodic oscillations, with each PPG peak occurring shortly after the corresponding R-wave peak. This is expected physiological behavior, supporting the suitability of this dataset for synchronicity-dependent analyses such as pulse transit time (PTT). It should be noted that PPG waveforms represent photodiode light intensity; inverting this signal produces the expected pressure-like morphology [35,52].

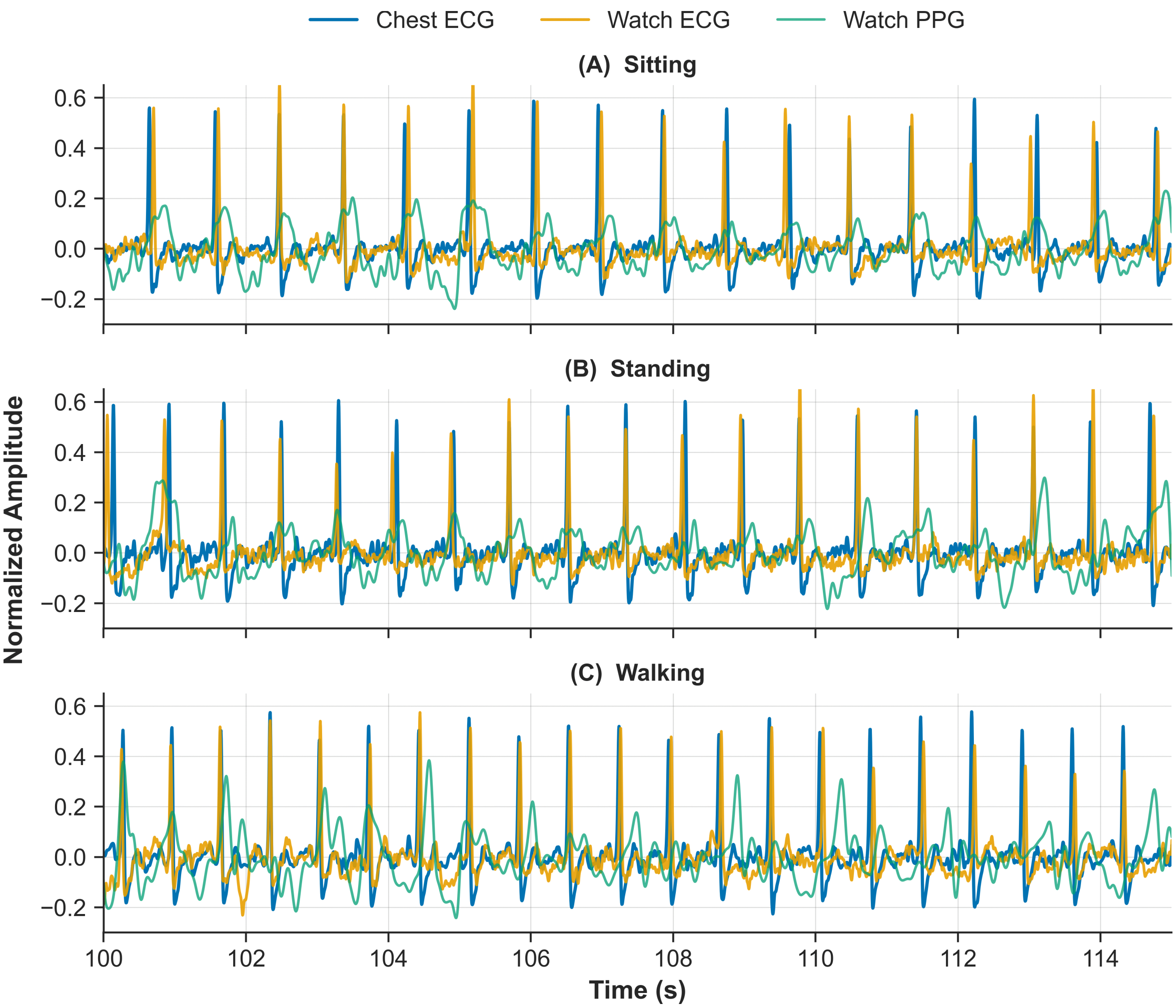

Fig 3. 15-second sample of ECG and PPG waveforms from participant p28 across (A) sitting, (B) standing, and (C) walking states.

The raw ACC signals for a representative subject are visualized in Fig 4 across physical states. The sitting and standing states show approximately constant signals, consistent with the expected behavior during static positions. While both states are relatively stable, there are noticeable differences in the X- and Z-axis values, likely due to the variations in sensor orientation between postures. In contrast, the walking state exhibits oscillatory waveforms across all axes due to rhythmic upper body motion during gait cycles. Although participants were required to maintain wrist ECG contact while walking, the ACC sensors still capture subtle upper arm swings. The distinct signal morphologies between static and dynamic states highlight the value of ACC signals for physical state-dependent filtration in cardiovascular analyses. Specifically, during repetitive dynamic movements such as walking, ACC signals can be leveraged to identify and remove motion-related frequencies from ECG and PPG signals while preserving cardiac-related frequencies such as HR.

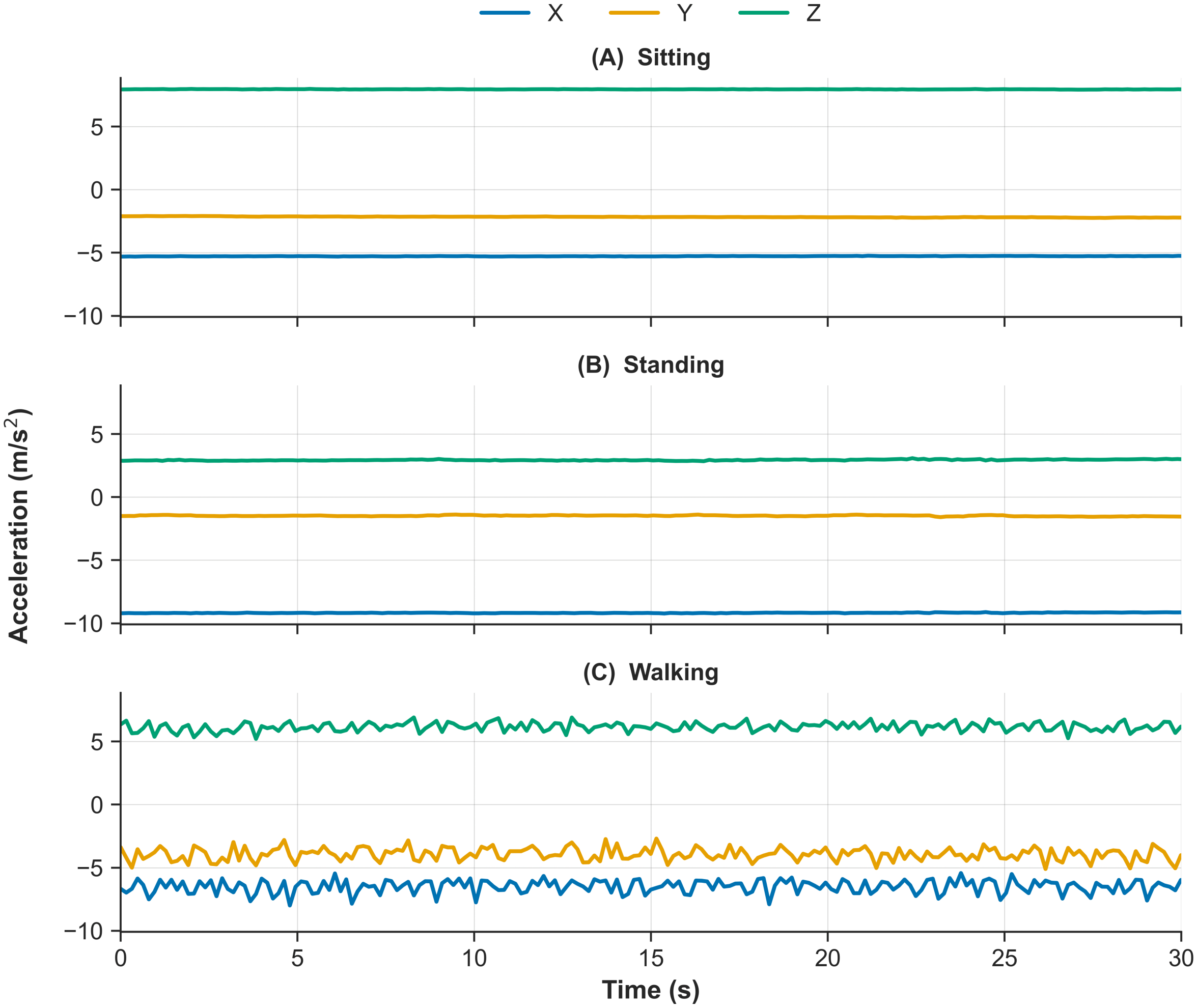

Fig 4. 30-second sample of triaxial ACC sensor waveforms from subject p16 across (A) sitting, (B) standing, and (C) walking states.

### 2.3 Sampling Rate Assessment

Verification of sampling stability is crucial in semi-naturalistic conditions, where device loads and system buffering can introduce timing deviations [36,53]. Following the validation method presented by Park et al. [36], the mean sampling rates of the raw smartwatch and chest ECG signals were computed for each physical state to verify that time intervals did not vary significantly during acquisitions. These time intervals were also computed during the five BP measurements. The calculated sampling rates are presented in Table 3. All biosignals were within 1Hz of their nominal rates, indicating stable sensor timing across subjects and recording sessions. Minor sampling rate fluctuations are expected, as internal sensor oscillators can exhibit small frequency variations [53].

Table 3. Mean sampling rates computed from raw sensor time intervals across recording sessions (SW: smartwatch).

| Recording Session | Chest ECG | SW ACC | SW ECG | SW PPG |
|---|---|---|---|---|
| Sitting | 250.13 Hz | 6.27 Hz | 250.93 Hz | 68.43 Hz |
| Standing | 250.14 Hz | 6.27 Hz | 250.95 Hz | 68.44 Hz |
| Walking | 250.11 Hz | 6.27 Hz | 250.96 Hz | 68.44 Hz |
| BP Measure 1 (seated) | 250.68 Hz | 6.27 Hz | 250.69 Hz | 68.36 Hz |
| BP Measure 2 (seated) | 250.80 Hz | 6.27 Hz | 250.75 Hz | 68.38 Hz |
| BP Measure 3 (standing) | 250.72 Hz | 6.27 Hz | 250.72 Hz | 68.37 Hz |
| BP Measure 4 (standing) | 250.71 Hz | 6.27 Hz | 250.74 Hz | 68.38 Hz |
| BP Measure 5 (standing) | 250.97 Hz | 6.27 Hz | 250.71 Hz | 68.37 Hz |

### 2.4 Signal Assessment

The strength of the signals across physical states were evaluated based on the formula presented by Park et al. [36] for classifying weak PPG and ECG signals. A segment is considered weak if the peak-to-peak values is less than 10% of the window's standard deviation.

$$\text{Weak PPG or ECG if: Peak-to-peak (segment)} < 0.1 * \sigma \text{ (window)} \qquad (1)$$

The 4-minute ECG recordings were first min-max normalized and zero-centered, then segmented into 30-second windows to calculate standard deviation. This window length was chosen to align with the typical smartwatch ECG duration used in analyses[41]. PPG recordings followed the same preprocessing, with the addition of NeuroKit2 filtering to remove baseline wander. Each 30-second window was further segmented into non-overlapping 2-second segments to compute peak-to-peak values. A window was considered weak if any of its 2-second segments matched the weak signal definition. As shown in Fig 5, windows with lead fall-offs or signal saturation were considered invalid and defined as having zero peak-to-peak values, as suggested by Liu et al. [54]. Windows were also invalid if they contained any not a number (NaN) values. For ACC signals, the criteria described by Park et al. [36] was implemented here, where windows with ACC vector magnitudes exceeding an upper bound were considered invalid. The upper bound was set to 1.5 g after reviewing the findings of Straczkiewicz et al. which indicate that ACC values during walking typically remain below this value [55]. ACC measurements, recorded in m/s$^2$, were converted to gravitational units before evaluating validity.

An additional assessment for ECG signal strength was introduced to identify windows with both strong and weak QRS complexes, such as in the such as in the example provided in Fig 5. Each 30-second window's peak-to-peak value was compared with the ones derived from its internal 2-second segments. A 2-second ECG segment was labelled weak if its peak-to-peak value was less than 50% of the 30-second window's peak-to-peak reference. To avoid overclassifying segments as weak, a 30-second window was marked as weak if more than 50% of its 2-second segments were classified as weak.

The results of this assessment are presented in Table 4. The smartwatch ECG was the only modality that contained weak or invalid windows, which was expected due to the dependence of signal quality on prolonged finger contact. The sitting and standing states have similar rates of 5.31% and 5.00% respectively, reflecting the stable conditions. The walking state showed the highest proportion of weak or invalid signals at 12.50%, illustrating the challenged of prolonged smartwatch ECG acquisition during physical activity. Notably, no NaN values were detected across modalities and physical states, demonstrating the benefit of the offline data transmission system.

Table 4. Percentage of weak or invalid 30-second segments across modalities and physical states (SW: smartwatch).

| Physical State | Chest ECG | SW ACC | SW ECG | SW PPG |
|---|---|---|---|---|
| Sitting | 0.00 % | 0.00 % | 5.31 % | 0.00 % |
| Standing | 0.00 % | 0.00 % | 5.00 % | 0.00 % |
| Walking | 0.00 % | 0.00 % | 12.50 % | 0.00 % |

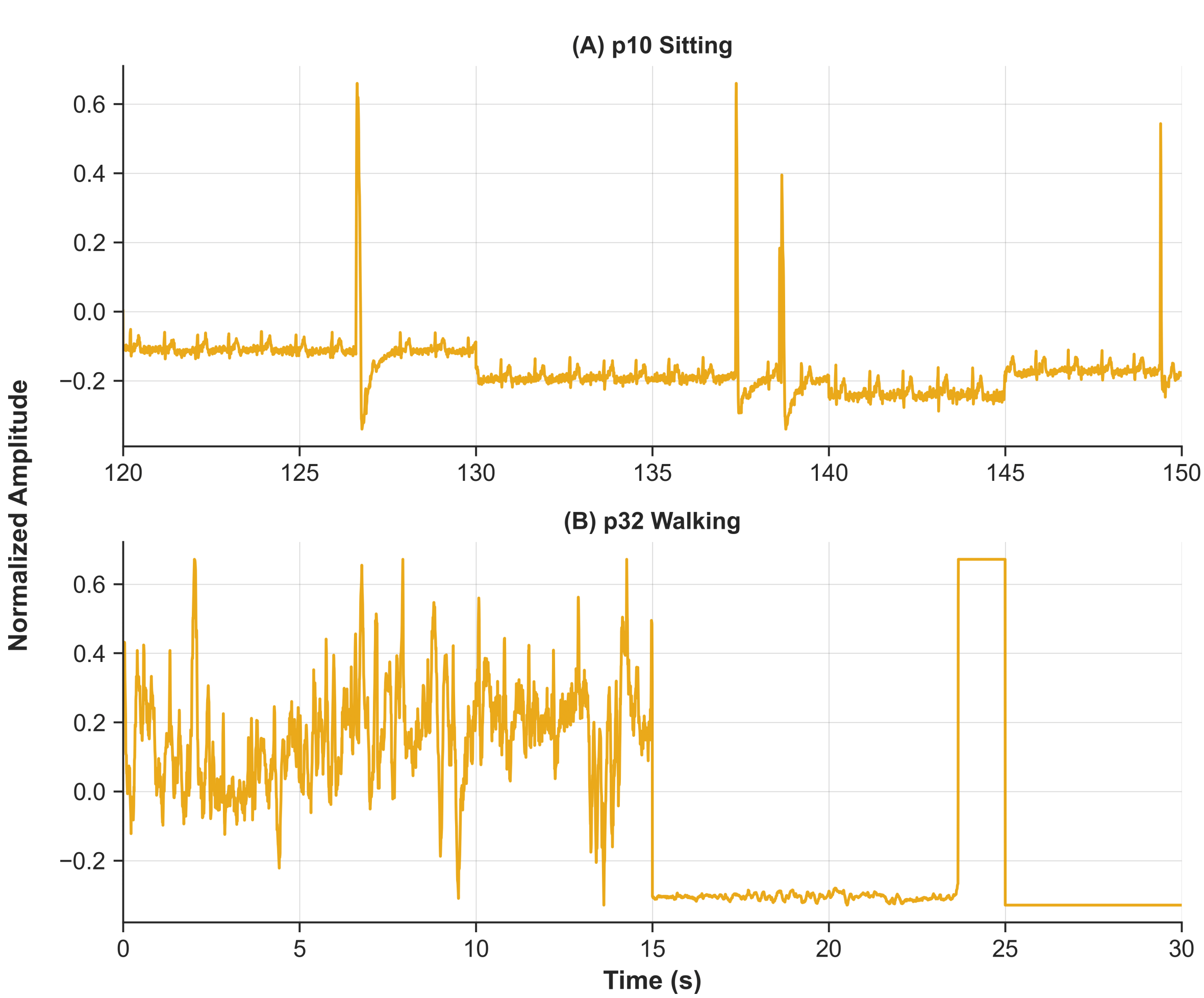

Fig 5. 30-second smartwatch ECG example of (A) a window with strong and weak QRS amplitudes, and (B) a window with signal saturation and/or lead fall-off issues leading to a square wave morphology.

### 2.5 Electrocardiogram Morphology Comparison - QRS and PR Intervals

To compare the signal morphology of the single-lead smartwatch ECG with the chest ECG, statistical analyses on the QRS and PR intervals were performed. The QRS interval is defined as the duration between the onset and offset of the QRS complex, and used as a predictor for cardiovascular mortality [56,57]. The PR interval is defined as the duration between the onset of the P-wave and the onset of the QRS complex, and is also a commonly-used biomarker for cardiovascular health [58,59].

Automatic ECG delineation was conducted using the Peak Prominence ECG Delineator, which applies physiological constraints to identify fiducial points with high temporal accuracy [60]. Following the methodology of Van Der Zande et al. [41], mean durations were calculated for each participant and activity state. Bland-Altman plots were then generated to compare bias and limits of agreement (LoAs) between the smartwatch and chest ECG measurements (see Fig 6). A paired *t*-test was conducted to evaluate the statistical significance of these biases across states, where $p < 0.05$ was considered statistically significant.

Minimal bias was observed for QRS intervals across all three conditions (sitting: +1.51 ms, standing: +1.04 ms, walking: +1.93 ms) in addition to narrow LoAs. During standing, the QRS bias was not statistically significant ($p = 0.085$). Although the sitting and walking state biases were statistically significant ($p < 0.05$), their magnitudes were small and corresponded to 1-2% of normal QRS intervals (80-100 ms). In contrast, smartwatch PR intervals exhibited a strong bias towards overestimation across physical states (sitting: +10.09 ms, standing: +13.22 ms, walking: +17.71 ms) with wide LoAs. These biases were statistically significant in all states ($p < 0.001$), suggesting that there is a consistent overestimation of PR intervals by the smartwatch ECG. Since this bias was not observed for QRS interval measurements, it is likely attributable to inaccuracies in P-waves detection, potentially due to attenuation of these features. This observation is consistence with previous reports on the limitations of Lead I smartwatch ECGs [61,62].

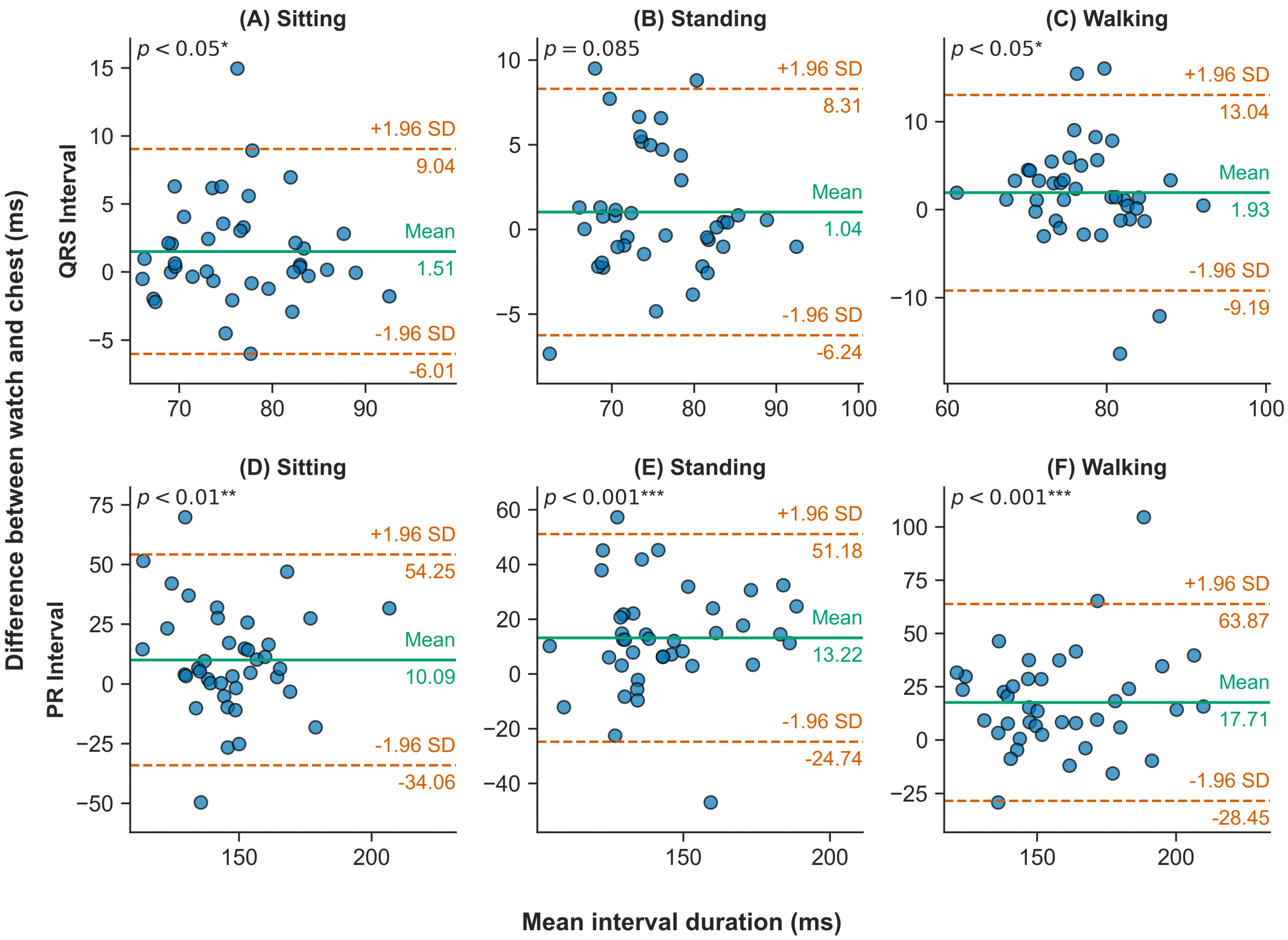

Fig 6. Bland-Altman plots for QRS intervals (A) – (C) and PR intervals (D) – (F) across physical states between smartwatch and chest ECG methods.

## 2.6 Heart Rate Analysis

The synchronization of the smartwatch signals enables applications such as long-term HR monitoring across daily activities. Fig 7 illustrates how instantaneous heart rate (IHR) can be computed from chest ECG, smartwatch ECG, and smartwatch PPG during motion using walking data. IHR is defined as 60 seconds divided by the duration, in seconds, between successive ECG R-peaks to yield a value in beats-per-minute (BPM) [63,64]. Smartwatch HR values were estimated using the adaptive notch-filtration approach proposed by Zheng et al. [65] which identifies candidate frequencies for notch filtering based on raw ACC signals. These candidate frequencies correspond to motion oscillations, as shown in Fig 4, unless there is overlap with ECG or PPG frequency peaks. In this work, IHR was computed using a 10-second sliding window, a 1-second step size, and a tolerance of 10 beats per minute (BPM).

Although the adaptive notch-filtration approach was originally designed for PPG signals during physical activity, the implementation also performs well with smartwatch ECG signals. Notably, the choice of ACC axis for this approach appears to have significant effects on the estimated smartwatch IHRs relative to the chest ECG reference. Future analyses may benefit from examining how careful selection or fusion of ACC axes can improve filtration techniques across the physical states and participant subgroups. This example highlights the value of synchronized ACC signals alongside wrist-based ECG and PPG, demonstrating how these modalities can be integrated to compare HR estimation techniques for continuous monitoring.

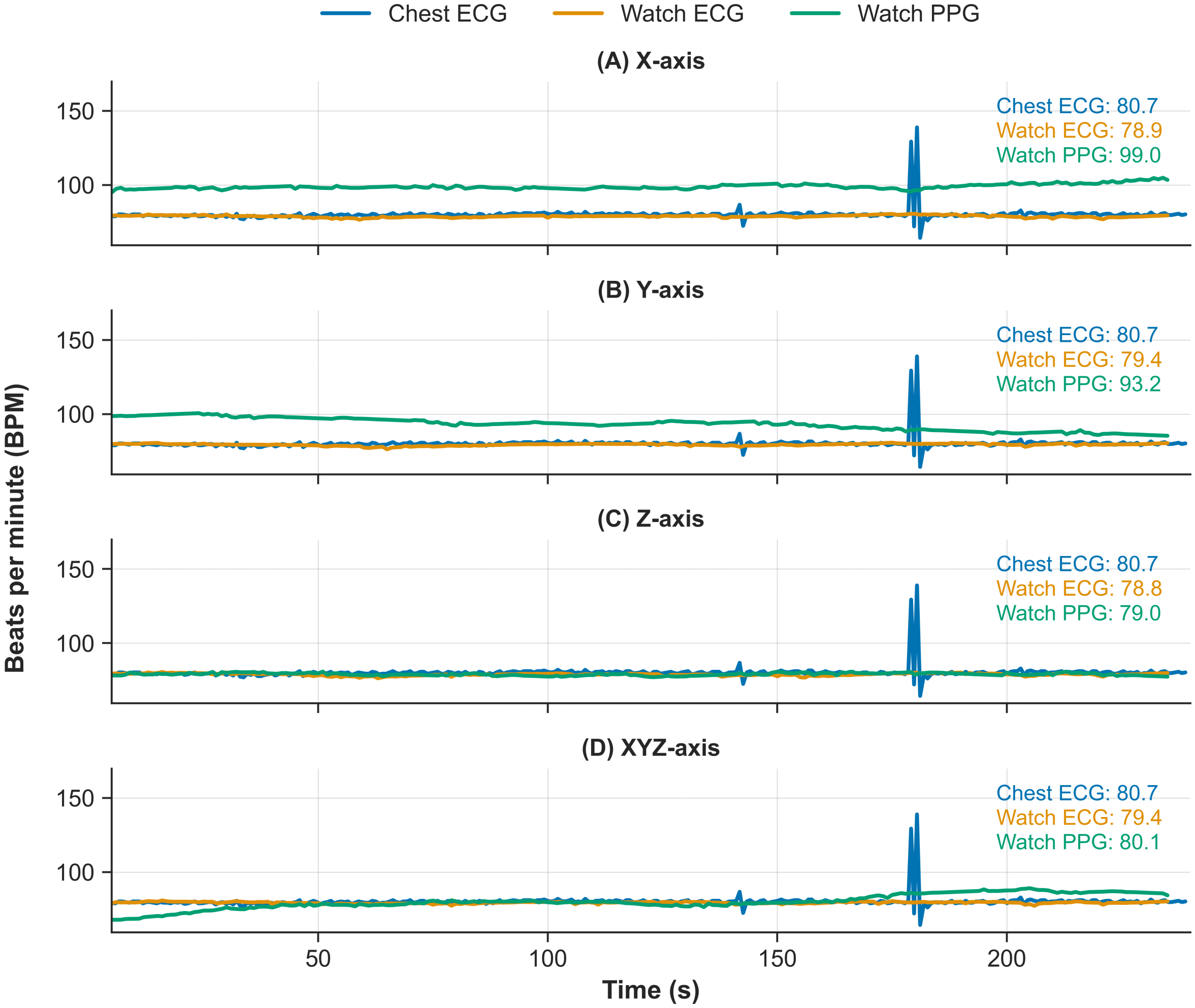

Fig 7. IHR estimation plots for participant p03 during walking using different ACC axes for adaptive notch-filtration on the smartwatch ECG and PPG signals.

### 2.7 Blood Pressure Measurements

Upper arm BP measurements were recorded for each participant, and the distributions of systolic and diastolic pressures are shown in Fig 8. These BP recordings can be analyzed alongside synchronized chest ECG and smartwatch ECG, PPG, and ACC signals collected during cuff acquisition. One potential application is cuffless BP estimation, which typically pairs distal PPG waveforms with chest ECG R-peaks to compute PTT [66]. While existing PTT-based approaches using smartwatches have primarily relied on wrist-based PPG combined chest ECG, the HEART-Watch dataset uniquely provides simultaneous wrist-based ECG and ACC signals, in addition to detailed demographic information. This multimodal combination may provide researchers with new avenues for continuous BP estimation techniques using consumer-grade smartwatches by incorporating complementary smartwatch-derived signals.

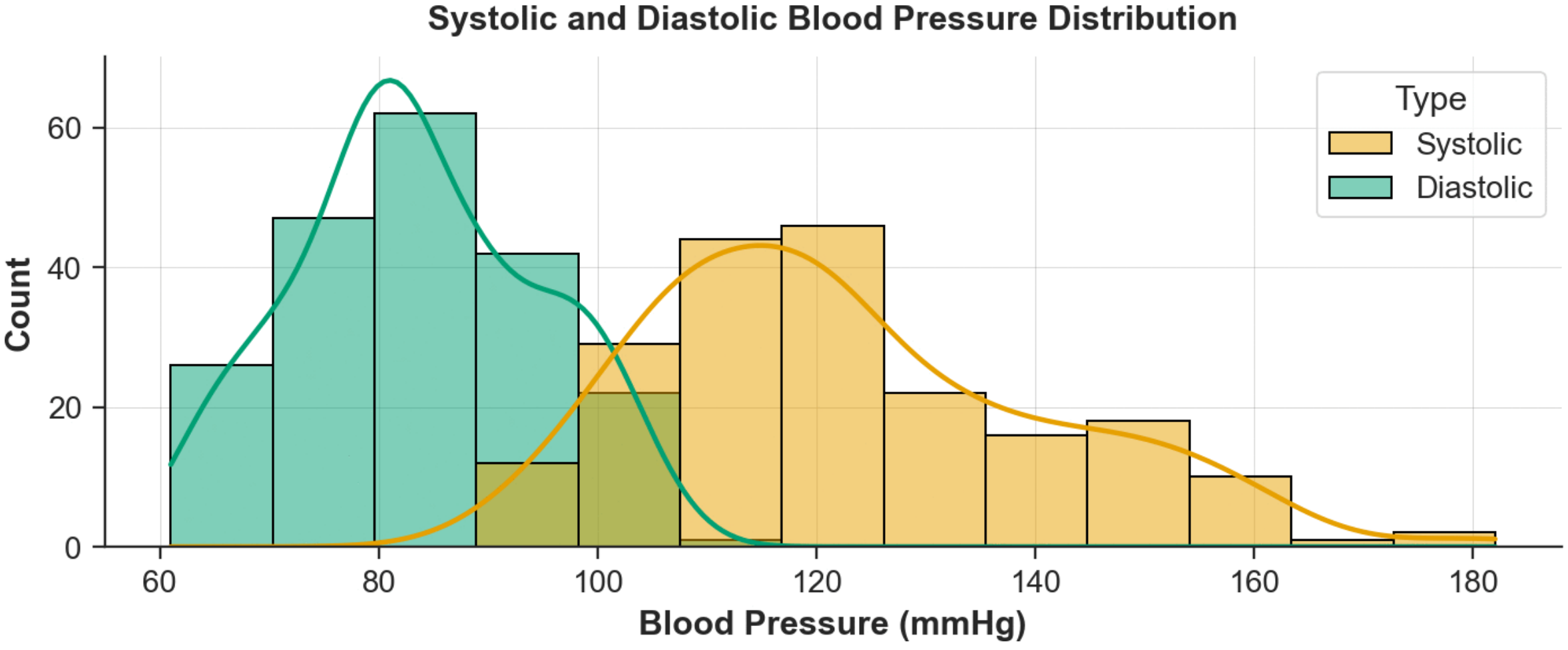

Fig 8. Histogram with overlaid kernel density estimates for systolic and diastolic cuff BP measurements.

## 3 DISCUSSION

HEART-Watch is unique from existing cardiovascular datasets in several aspects. A key advantage of this dataset is the direct use of a consumer-grade smartwatch for data collection, as these devices are often less accurate than medical-grade devices and are not optimized for biosignal acquisition. This approach ensures that algorithms are trained and evaluated on signals that capture the limitations of consumer-grade devices, resulting in performance that is more representative of real-world deployment scenarios on similar hardware. The availability of raw ECG, PPG, and ACC signals affected by real-world noise sources (e.g., motion, posture, poor wrist contact) allows researchers to implement their own signal processing pipelines for common aggregate measures such as HR and HRV – something that is not possible with many existing datasets that only provide proprietary algorithm outputs. HEART-Watch also includes extended smartwatch ECG recordings of up to 4 minutes across multiple physical states, which, to the best of our knowledge, has not been previously reported for consumer-grade devices. The multimodal structure, with synchronized chest ECG and cuff BP measures, supports fusion-based approaches for tasks such as cuffless BP estimation under realistic wearable conditions that are not typically captured in strictly controlled clinical datasets. HEART-Watch also includes a demographically diverse adult cohort (n = 40) across eight racial groups and ages up to 75 years, with detailed participant metadata to facilitate subgroup fairness assessments. Although the sample size is not intended for the training large-scale machine learning (ML) models, it provides a valuable resource for benchmarking algorithm performance and identifying potential biases that are often difficult to assess using existing datasets with limited diversity.

This study was conducted in a semi-naturalistic environment that relied on participant compliance for extended periods of time. During sitting and standing phases, participants were permitted reasonable movement rather than being strictly static across 4 minutes, which may have introduced noise into the acquired signals. Participants also showed varying levels of compatibility with both the chest ECG system and the smartwatch for achieving high-quality waveforms. During chest ECG calibrations before each recording, females often exhibited higher levels of noise present in their signals, a finding that aligns with previous research suggesting that breast tissue may interfere with signal quality [39]. Similarly, our team visually noted degradations in chest ECG quality, specifically for QRS complex identification, when participants had rounded shoulders while collecting smartwatch ECGs. These artifacts may be related to changes in the heart's orientation within the chest cavity, and deviations from traditional supine recordings have previously been associated with significant variations in ECG morphology [67]. As a result, reference chest ECG signals likely contain noise or motion artifacts under to these conditions.

Participants were also instructed to maintain contact with the smartwatch using their right index finger during smartwatch ECG acquisition. It is likely that some participants did not maintain optimal contact for the entire duration of the recordings, leading to saturation or motion artifacts like those presented in Fig 5. All recordings are included in HEART-Watch, as these phenomena are encouraged as robustness concepts for ML methods in healthcare [50]. Similarly, the wrist-based PPG waveforms are susceptible to artifacts from skin tone, wrist contact, skin perfusion, and motion [21]. Given the diverse participant cohort and range of physical states, many PPG signals will likely deviate from ideal waveform morphology, and future analyses may benefit from incorporating the first PPG channel as a reference for signal correction. The low sampling rate of the ACC data (6 Hz) is also a limitation of this dataset, as it is insufficient for precise motion analyses such as gait or activity characterization, which typically requires a minimum of sampling rate of 32 Hz [68]. However, as shown in sections 2.2 and 2.6, this sampling rate is sufficient for capturing physical state transitions and providing frequency-related information for reducing motion-artifacts in the ECG and PPG signals.

Finally, the use of separate devices and system clocks introduced the risk of clock drift and jitter, which are common challenged in clinical recording environments [21,53]. To minimize these concerns, smartwatch clocks were manually synchronized with Coordinated Universal Time before each session, and researchers avoided recording with watch battery levels below 50% to limit system lags or sensor instability.

Wearable technology is rapidly advancing as the market for personalized healthcare grows on a global scale. Consumer-grade smartwatches, in particular, provide accessible and affordable continuous health monitoring to the general public. However, the development of robust signal processing and AI algorithms requires training and validation with realistic, long-form raw sensor data collected across diverse participants. HEART-Watch addresses this need by providing synchronized smartwatch signals – including prolonged ECG, PPG, and ACC recordings – alongside chest ECG and BP measures across common physical states. Although the semi-naturalistic nature of this dataset may introduce artifacts that can degrade signal quality, it also reflects the practical challenges of wearable sensing. Rather than excluding such data, it can be leveraged to design and validate algorithms that are resilient to these variations. In this way, HEART-Watch offers the research community a resource for advancing the development of consumer-grade smartwatch algorithms that are both robust and equitable.

# 4 MATERIALS AND METHODS

## 4.1 Ethics & Recruitment

This study was reviewed and approved by the University Health Network (UHN) Research Ethics Board (UHN REB #24-5559) on August 2nd, 2024. Recruitment was conducted through the KITE Research Institute, UHN's Study Recruitment webpage open to the public [69], on social media, and via physical posters at several UHN-affiliated hospitals. Participants were recruited if they were 18 years or older, able to perform light physical activities without assistance, had intact skin at sensor placement sites, and no self-reported history of disease that could affect breathing or HR. The inclusion and exclusion criteria were designed to recruit a diverse set of healthy participants that did not have morbidities that may confound cardiovascular biomarkers. Participants provided written informed consent to the release of their de-identified data through a control-accessed database prior to data collection.

### 4.2 Study Setting

Participants were scheduled for a single 1-hour visit to the Challenging Environment Assessment Laboratory (CEAL) within the Toronto Rehabilitation Institute. This controlled, indoor setting was chosen to standardize environmental conditions across sessions such as ambient lighting, sound, and temperature which could influence biosignal acquisition. The duration of the session was chosen to accommodate all protocol tasks within a single visit, balancing participant burden and data completeness. Although repeated sessions were outside the scope of this work, examining inter-session variability of any significant findings would be a valuable direction for future research.

### 4.3 Data Acquisition Devices

Separate devices were used to collect smartwatch signals, reference chest ECG, and BP values respectively. Since there is no common point in the system for data synchronization, Unix timestamps were generated for each sensor event and stored during collection. Smartwatch timestamps and chest ECG timestamps were recorded with nanosecond and millisecond precision respectively. Hardware details are summarized in Table 5. Smartwatch sampling rates are reported as observed at each sensor's default setting during preliminary trials; maximum smartwatch rates were not enabled because prolonged, synchronized recordings exhibited sensor instability (e.g. information loss). The default rate of the Olimex shield is 256 Hz but was set to match the sampling rate from the smartwatch ECG to 250 Hz for consistency between ECG devices.

Table 5. Summary of hardware for biodata acquisition.

| Device | Signal Acquisition | Modality | Sampling rate | Data Transfer |
|---|---|---|---|---|
| Google Pixel Watch (2022) | AFE4950, 3-axis accelerometer | ECG, PPG, ACC | 250 Hz, 68 Hz, 6 Hz | Android Wireless Debugging |
| Olimex SHIELD-EKG-EMG, Arduino Uno | 3M Red Dot ECG Electrode (2249-50) | Chest ECG | 250 Hz | Micro-USB |

| Device | Signal Acquisition | Modality | Sampling rate | Data Transfer |
|---|---|---|---|---|
| Omron BP5000 Upper Arm Monitor | — | Cuff BP | — | Manual recording |

Modalities should be considered smartwatch-derived unless indicated. A '—' indicates that the category is not applicable.

### 4.4 Data Collection

The data collection procedure involves three main physical states: sitting, standing, and walking. BP measurements were recorded during the transition period between each of these states using the Omron digital monitor. Each participant was instructed to secure the watch to their left wrist independently to mimic real-world behavior, while the BP cuff was placed on their upper right bicep to avoid contaminating watch signals during cuff inflation and deflation [70,71]. Three gel electrodes were placed on the chest and torso to acquire reference chest ECG data, and the presence of QRS complexes were visually verified before each physical state and BP recording. Participants were instructed to place their right index finger on the crown of the smartwatch as the researchers initiated each recording. The participants followed this procedure for both the physical state and the BP recording sessions; however, physical state sessions were 4 minutes in duration while the BP sessions were stopped after the monitor reported systolic and diastolic readings. Four minutes was chosen as the duration for physical states as multiple studies have shown that this is the minimum duration required for common frequency-domain cardiovascular metrics, such as HRV low and very low frequency power in static and non-static physical states [72,73]. Due to the aforementioned physical constraints with collecting smartwatch ECG, this duration was not extended to minimize participant burdens during the walking portion of the procedure.

During BP recordings, cuff inflation was initiated after approximately 30 seconds of biosignal recordings to establish a non-perturbed baseline. This short pre-inflation baseline was acquired due to previous findings showing that distal blood flow is modulated during cuff inflation and deflation, which could affect PTT analyses if distal PPG waveforms were impacted [74,75].

Participants were asked to avoid speaking or excessive motion during data acquisition, however, with the goal of semi-naturalistic data acquisition in mind, they were permitted to do so within reason. Additionally, smartwatch data was not accessed or visualized during data collection in real-time due to technical limitations with the device and security restrictions. The standing and walking recordings were conducted on the treadmill, Norditrack C1750, with the treadmill speed set to a moderate walking pace unique to each participant's comfortability. The maximum speed was set to 4.83 km/h (3.0 mph) for participant safety in accordance with our ethics approvals. To avoid variable delays and missing packets from wireless transmission [53], an offline system was designed such that the chest data was streamed serially through USB to the research laptop and the watch data was stored locally on the device. Data files, stored as CSVs, were manually extracted from the watch through Android Studio Wireless Debugging and stored on the research laptop after each participant completed the procedure. The deidentified CSVs were manually uploaded to UHN OneDrive platform after each visit. A visualization of a standard visit is presented in Fig 9.

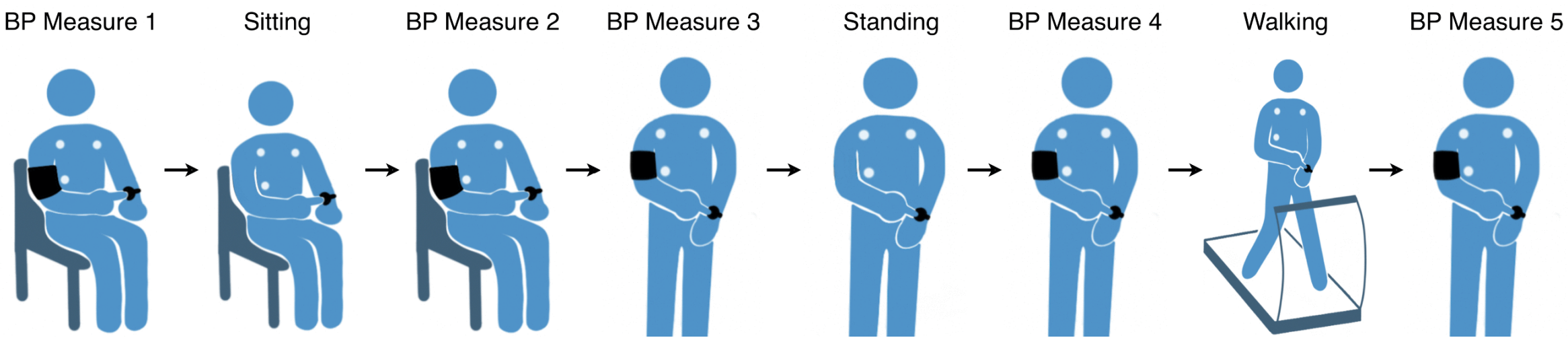

Fig 9. Visualization of data collection procedure: sitting BP → 4 minutes sitting → seated BP → standing BP → 4 minutes standing → standing BP → 4 minutes walking → standing BP.

### 4.5 Data Organization

The raw and synchronized versions of the data have been provided as CSVs within HEART-Watch. The raw version of the data refers to the smartwatch and chest data that were recorded in their original sampling rates (refer to Table 5) and stored as separate CSVs as shown in Fig 10. The synchronized version compiles the raw files into a single CSV by zero-centering signals, applying linear interpolation to 250 Hz, and aligning via Unix timestamps. Linear interpolation to 250 Hz was chosen to match the maximum native sampling rate of the chest and smartwatch ECG. This resampling technique constructs intermediate data points using linear polynomials without introducing new physiological features, though it may smooth high-frequency components. The raw CSVs are provided at each sensor's native sampling rate to allow researchers to apply alternative resampling strategies. ACC signals were not zero-centered to preserve the influence of gravity. Additional synchronization was performed by estimating a constant time domain shift based on mean R-peak time differences between the smartwatch and chest ECG data and shifting the watch signals accordingly to account for potential clock drift. The synchronized CSVs do not have any normalization or filtering techniques applied to the data to provide flexibility in processing pipelines for researchers. The following sections describe the physiological background and data-acquisition procedures for each modality.

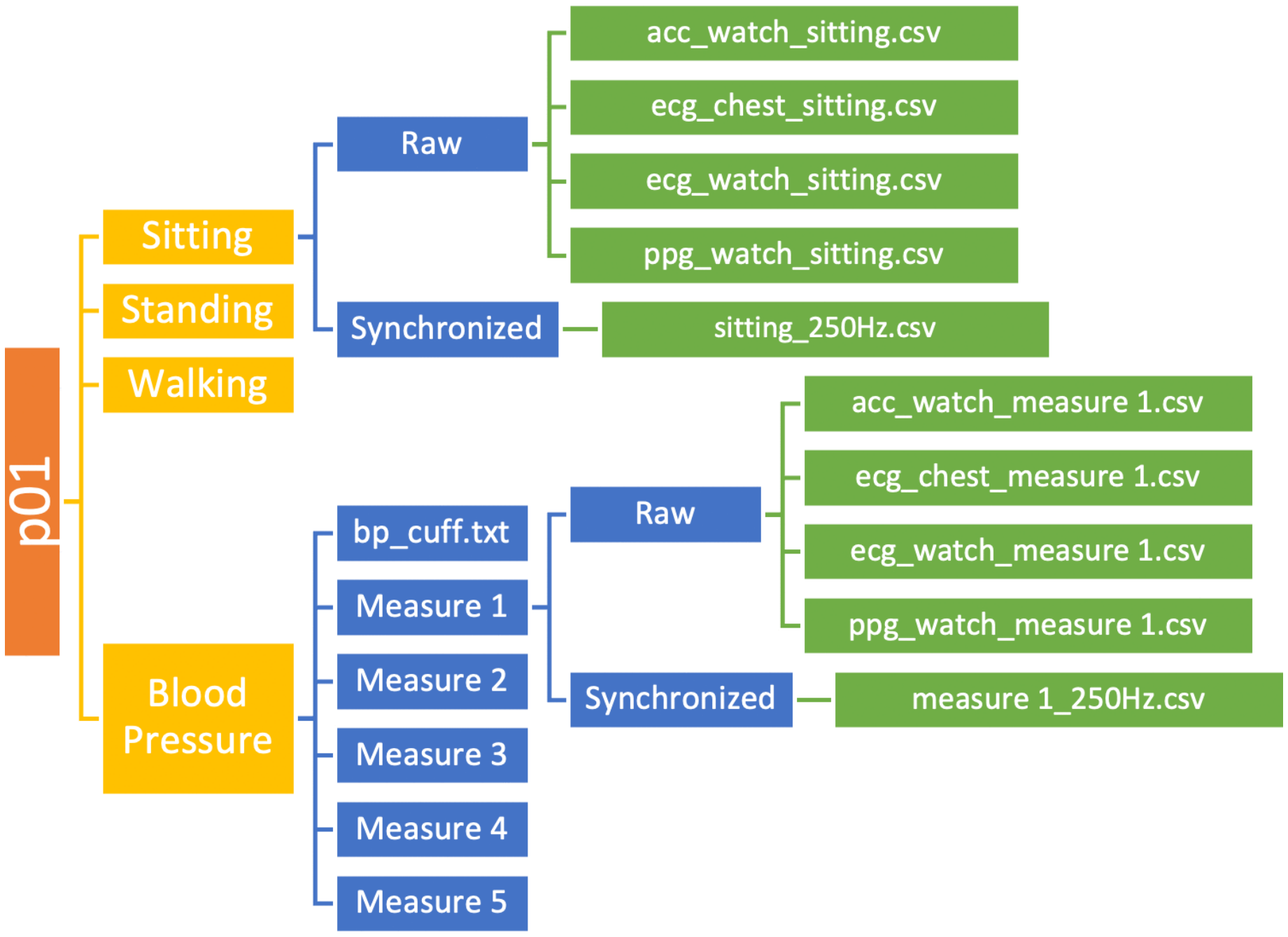

Fig 10. A visualization of the directory structure of HEART-Watch for raw and synchronized versions of the dataset for one participant.

*4.5.1 Chest Electrocardiogram*

The reference chest ECG used in this study is a bipolar Lead I configuration consistent with Einthoven's triangle [43]. A modified 3-electrode system was employed, with left and right electrodes placed at the respective infraclavicular fossae, and the ground electrode at the right lower rib cage [76]. The clavicle and rib cage bones were used as anatomical landmarks to guide electrode placement for consistency across participants. This electrode positioning, motivated by the work of Mason and Likar for exercise ECG [77], has been shown to produce similar Lead I data to traditional 12-lead ECG systems [78] , and was chosen over limb placements to avoid excessive motion artifacts from smartwatch use or during walking. The gel electrodes, 3M Red Dot 2249-50, were chosen for their low impedance and stable performance among commercially available pregelled skin electrodes [79]. The Arduino Uno and Olimex shield were also chosen as a low-cost ECG monitoring system with strong signal amplitude due to the chain of onboard filters that increase total gain [80].

Sensor values streamed from this board represent the potential difference between the positive left electrode and the negative right electrode. A diagram illustrating the chest ECG setup is shown in Fig 11.

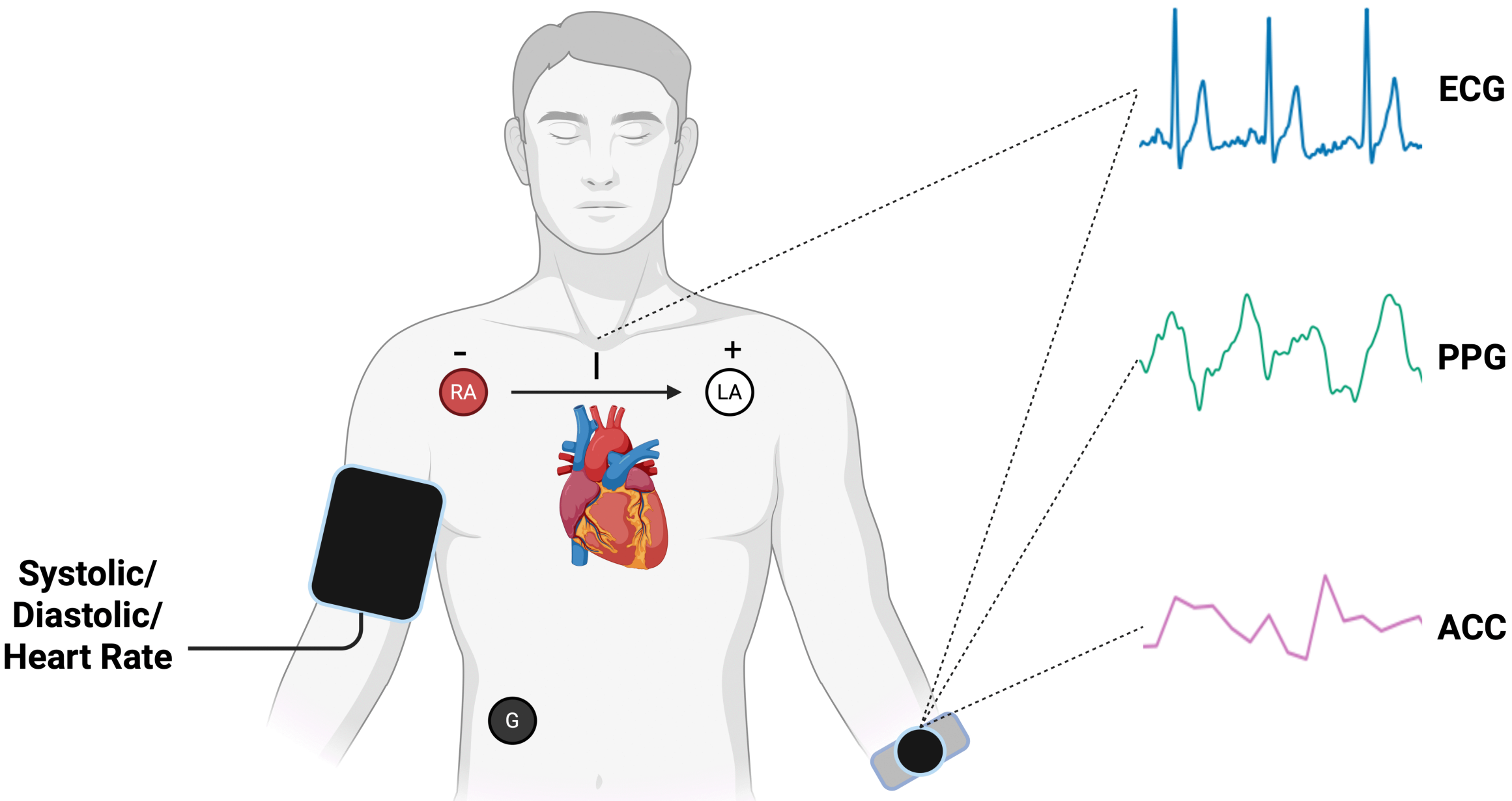

Fig 11. The data acquisition setup using a Lead I setup for chest ECG, Google Pixel Watch on the left wrist for ECG, PPG, and ACC signals, and a BP cuff on the right bicep for systolic and diastolic BP measurements. Created in BioRender. Kaetheeswaran, J. (2025) https://BioRender.com/k7cdwwt

*4.5.2 Smartwatch Electrocardiogram*

Consumer-grade smartwatch ECGs are also bipolar Lead I recordings. The potential difference is computed between the right index finger touching the crown of the watch (negative) and the electrode on the backside of the watch contacting the wrist (positive) [43]. In this way, the Lead I signals generated from the chest ECG can be directly compared with the signals from the smartwatch ECG. ECGs are useful for providing direct insights into cardiac functions through the analysis of each PQRST complex, which is a well-defined pattern representing the depolarization and repolarization of the heart for each cardiac cycle [81]. The frequency of PQRST waves can be used to calculate HR and identify irregular durations or deflections that could represent cardiac arrythmias such as AF [82].

#### *4.5.3 Smartwatch Photoplethysmography*

The smartwatch PPG sensor is typically composed of a light emitting diode (LED) and a photodetector [83]. The LED, often a green LED in wrist-based wearables, emits light through the wrist tissue, while the photodetector measures the reflected light [84]. The propagation of light energy through the wrist tissue is based on Beer-Lambert law, with the changes measured by the photodetector corresponding to changes in blood volume due to cardiac cycles [85]. In contrast to watch-based ECGs, smartwatch PPGs do not require active contact with the contralateral hand, allowing for passive, continuous monitoring from normal wear [42]. For this reason, PPG is the primary signal used for cardiovascular monitoring in wearable devices, as it can be used to estimate HR, blood oxygenation, sleep stages, and cuffless BP [6]. The AFE4950 hardware in the Google Pixel Watch contains dual receiving channels: a primary PPG channel and a reference channel for motion or ambient light correction [86,87]. Our team reached out to the AFE4950 manufacturer, Texas Instruments, to request a full datasheet to identify each channel's exact purpose but were not granted access. Based on our internal testing, the second PPG channel, denoted *ppg_watch_two*, should be used for pulse waveform analysis as it contains the pulsatile portion of traditional PPG signals. Therefore, it is likely that the first PPG channel, denoted *ppg_watch_one* is a reference channel for signal recovery, so it has been included in our published dataset but not utilized in the analyses and visualizations presented in this work.

#### *4.5.4 Smartwatch Accelerometer*

The electrical signals outputted from an ACC sensor are generated in response to changes in the body's velocity per second [88]. Smartwatches are often equipped with three-axis ACC sensors that can capture dynamic movements of the device. ACC signals have been widely incorporated into existing smartwatch monitoring algorithms as it can be measured passively like PPG, and can provide useful estimations of activity metrics such as step count [6]. The utility of this modality for cardiovascular monitoring can be extended by providing valuable event-related information for filtration of motion artifacts from smartwatch ECG and PPG signals [21].

#### *4.5.5 Inflatable BP Cuff*

Upper arm BP monitors estimate systolic and diastolic values by applying pressure on the brachial artery and gradually deflating the cuff while measuring pressure oscillations [89]. The Omron BP5000, a US FDA-registered medical device, was chosen to provide accurate readings while minimizing motion artifacts [90]. The five readouts from the digital monitor for each participant are available in the format *systolic/diastolic/heart rate*. The first two measurements were taken while seated, while the remaining three readings are from a standing position. BP measurements could not be obtained while walking because the BP5000 device automatically terminates readings when motion artifacts are detected [90].

### 4.6 Smartwatch Application

The raw sensor values from the Google Pixel Watch were collected through a custom application developed with Android Studio. The SensorManager API provided direct access to the ECG, PPG, and ACC SensorEvents which were recorded and stored in an on-device Room database along with Unix timestamps [91,92]. Kotlin coroutines were programmed to ensure that sensor database inserts were performed asynchronously on background threads without affecting ongoing sensor readouts and app performance [93].

### 4.7 Dataset and Code Availability

The HEART-Watch dataset is maintained through the UHN OneDrive platform. Researchers who are interested in obtaining this dataset for research purposes should contact the co-principal investigator at milad.lankarany@uhn.ca. Further queries can be made to the UHN REB (reb@uhn.ca) citing study 24-5559. Data will be made available to researchers who submit details of a study plan and sign the data access agreement to maintain participant confidentiality and ensure compliance with ethical standards. The Jupyter Notebook code for data organization and validation will be shared alongside the HEART-Watch dataset following successful requests. An outline of the available scripts is described here:

(i) *synchronize.ipynb* – Compiles raw CSVs into a single temporally synchronized CSV for each subject with all modalities at 250 Hz following the process described in section 4.5.

(ii) *demographics.ipynb* – Generates plots for participant demographics as shown in Fig 1 and Fig 2 . Also generates the histogram in Fig 8 to visualize blood pressure distributions.

(iii) *visualize.ipynb* – Generates subject-specific time-series visualizations of ECG, PPG, and ACC signals as shown in Fig 3 and Fig 4.

(iv) *sampling_rate.ipynb* – Computes mean sampling rates of ECG, PPG, and ACC signals across all participants and physical states as shown in Table 3.

(v) *signal_strength.ipynb* – Computes percentage of 30-second segments containing weak or invalid ECG, PPG and ACC signals across all participants and physical states as shown in Table 4.

(vi) *ecg_morphology.ipynb* – Computes QRS and PR intervals using the chest and smartwatch ECGs, and formats the results into the Bland-Altman plots shown in Fig 6.4.54.5

(vii) *heart_rate.ipynb* – Calculates heart rate for a subject from ECG and PPG signals after adaptive filtration is applied using the ACC signal, and generates plots for each ACC axis as shown in Fig 7.4.5

**ACKNOWLEDGMENTS**

The authors would like to thank the team at the CEAL and our industry partner Wellspring Data Limited for their support in carrying out this work.